\documentclass[sigconf]{acmart}

\AtBeginDocument{%
  \providecommand\BibTeX{{%
    \normalfont B\kern-0.5em{\scshape i\kern-0.25em b}\kern-0.8em\TeX}}}

\setcopyright{acmlicensed}
\copyrightyear{2018}
\acmYear{2018}
\acmDOI{XXXXXXX.XXXXXXX}

\acmConference[Conference acronym 'XX]{Make sure to enter the correct
  conference title from your rights confirmation emai}{June 03--05,
  2018}{Woodstock, NY}
\acmISBN{978-1-4503-XXXX-X/18/06}

\definecolor{correct}{RGB}{70, 166, 92}
\definecolor{incorrect}{RGB}{195, 37, 37}
\definecolor{hint}{RGB}{191, 155, 31}

\usepackage{enumitem} 

\copyrightyear{2024}
\acmYear{2024}
\setcopyright{rightsretained}
\acmConference[AVI 2024]{International Conference on Advanced Visual Interfaces 2024}{June 3--7, 2024}{Arenzano, Genoa, Italy}
\acmBooktitle{International Conference on Advanced Visual Interfaces 2024 (AVI 2024), June 3--7, 2024, Arenzano, Genoa, Italy}\acmDOI{10.1145/3656650.3656667}
\acmISBN{979-8-4007-1764-2/24/06}

\begin{document}

\title{Visualizing Intelligent Tutor Interactions for Responsive Pedagogy}

\author{Grace Guo}
\authornote{Both authors contributed equally to this research.}
\orcid{0000-0001-8733-6268}
\affiliation{%
  \institution{Georgia Institute of Technology}
  \streetaddress{North Ave NW}
  \city{Atlanta}
  \state{Georgia}
  \country{USA}
  \postcode{30332}
}
\email{gguo31@gatech.edu}

\author{Aishwarya Mudgal Sunil Kumar}
\authornotemark[1]
\orcid{0009-0001-0793-8347}
\affiliation{%
  \institution{Georgia Institute of Technology}
  \streetaddress{North Ave NW}
  \city{Atlanta}
  \state{Georgia}
  \country{USA}
  \postcode{30332}
}
\email{amsk3@gatech.edu}

\author{Adit Gupta}
\orcid{0000-0002-1218-0630}
\affiliation{%
  \institution{Drexel University}
  \streetaddress{3230 Market Street}
  \city{Philadelphia}
  \country{USA}}
\email{adit.gupta@drexel.edu}

\author{Adam Coscia}
\orcid{0000-0002-0429-9295}
\affiliation{%
  \institution{Georgia Institute of Technology}
  \streetaddress{North Ave NW}
  \city{Atlanta}
  \state{Georgia}
  \country{USA}
  \postcode{30332}
}
\email{acoscia6@gatech.edu}

\author{Christopher J. MacLellan}
\orcid{0000-0003-3084-5189}
\affiliation{%
  \institution{Georgia Institute of Technology}
  \streetaddress{North Ave NW}
  \city{Atlanta}
  \state{Georgia}
  \country{USA}
  \postcode{30332}
}
\email{cmaclell@gatech.edu}

\author{Alex Endert}
\orcid{0000-0002-6914-610X}
\affiliation{%
  \institution{Georgia Institute of Technology}
  \streetaddress{North Ave NW}
  \city{Atlanta}
  \state{Georgia}
  \country{USA}
  \postcode{30332}
}
\email{endert@gatech.edu}

\newcommand{\grace}[1]{\textcolor{purple}{#1}}
\newcommand{\alex}[1]{\textcolor{teal}{#1}}
\newcommand{\adam}[1]{\textcolor{blue}{#1}}
\newcommand{\adit}[1]{\textcolor{green}{#1}}

\renewcommand{\shortauthors}{Guo and Kumar, et al.}

\begin{abstract}
  Intelligent tutoring systems leverage AI models of expert learning and student knowledge to deliver personalized tutoring to students.
  While these intelligent tutors have demonstrated improved student learning outcomes, it is still unclear how teachers might integrate them into curriculum and course planning to support responsive pedagogy.
  In this paper, we conducted a design study with five teachers who have deployed Apprentice Tutors, an intelligent tutoring platform, in their classes.
  We characterized their challenges around analyzing student interaction data from intelligent tutoring systems and built VisTA (Visualizations for Tutor Analytics), a visual analytics system that shows detailed provenance data across multiple coordinated views.
  We evaluated VisTA with the same five teachers, and found that the visualizations helped them better interpret intelligent tutor data, gain insights into student problem-solving provenance, and decide on necessary follow-up actions -- such as providing students with further support or reviewing skills in the classroom.
  Finally, we discuss potential extensions of VisTA into sequence query and detection, as well as the potential for the visualizations to be useful for encouraging self-directed learning in students.
\end{abstract}




\received{20 February 2007}
\received[revised]{12 March 2009}
\received[accepted]{5 June 2009}

\maketitle

\section{Introduction}


Intelligent tutoring systems \cite{koedinger2006cognitive} are a type of computerized educational technology that provides students with individualized practice problems, correctness feedback, and hints.
Leveraging AI models of expert knowledge and student learning, intelligent tutors estimate student knowledge from their interactions with the system to deliver personalized instruction.
While they have been widely demonstrated to improve student outcomes \cite{Pane1967,koedinger1997,Aleven2009}, few studies have looked at how teachers might be supported in better integrating intelligent tutoring systems into their curriculum and course planning in ways that enable responsive pedagogical practice \cite{Ladson-Billings:2014:CulturalPedagogy2.0, Morrison:2008:OperationalizingCulturalPedagogy, Paris:2012:CulturalPedagogy, Rosebery:2010:HeterogeneityLearning}.

Closely related to intelligent tutoring systems are other online learning technologies such as learning management systems (LMSs) and massive open online courses (MOOCs).
In prior studies, researchers have looked at how these online learning technologies might be supported through visualizations \cite{emmons2017mooc, dewan2021review, zhang2022towards}.
These visualizations, such as learner trajectories, help analyze how students navigate through course content \cite{goulden2019ccvis, lundqvist2019visualising}, and have been advocated as a useful means to both guide student-centered teaching \cite{sztajn2012learning, wilson2015teachers} and optimize online course design \cite{ginda2019visualizing}.

Specific to intelligent tutoring systems, other studies have looked at how visualization dashboards might be developed to help teachers analyze student data \cite{aleven2010developing, holstein2010luna, xhakaj2017effects, aleven2022dashboard, hou2022design, shanabrook2012visualization}.
However, the majority of these dashboards were focused on aggregate measures.
In contrast, the data collected by intelligent tutors are often fine-grained logs of student interactions and estimated skill mastery as they work through problems in the tutor.
There is thus potential for visualizations of problem solving trajectory (i.e. the history or ``provenance'' \cite{ragan2015characterizing} of steps taken to solve a problem) to help teachers better understand how students use intelligent tutors.
These insights can in turn allow teachers to better integrate tutors into their curriculum, and tailor their responses to individual student needs during classroom instruction.

In this paper, we leverage the fine-grained interaction logs collected by intelligent tutors to address new design opportunities structured around three central questions: 1) How can we visualize the provenance of student problem solving processes and overall performance? 2) How can we help teachers understand and operationalize this data? 3) How might teachers adjust classroom instruction to respond to individual student needs?

We work with Apprentice Tutors, a web-based intelligent tutoring platform currently deployed in university classroom settings.
Apprentice Tutors collects data on how students work through problems, their interactions, estimated skill mastery, correctness, and the time spent solving problems.
Collaborating with five teachers at a local state university who have deployed the tutor to their students, we conducted a design study \cite{sedlmair2012design} to introduce them to the data collected and understand their user tasks and goals.
Based on participant feedback, we identified three user challenges centered around: summarizing student engagement, identifying issues with each problem-solving step, and comparing student performance over time.
These challenges then guided our design of the VisTA visual analytics tool.

VisTA has four main views: the Overview, Student view, Problem Type view, and Details view.
Collectively, these views enable teachers to see aggregate metrics of student engagement, identify student problem solving strategies, compare problem solving provenance between different problem types and students, and identify skills students commonly struggle with.
We conducted a qualitative evaluation of our system with the same five teachers to assess its usability, effectiveness, and the insights generated.
From their feedback and suggestions, we found that VisTA helped teachers better understand student problem solving processes and decide on the necessary follow-up actions -- such as identifying when students require further support or which skills should be reviewed in the classroom.
Finally, we also discuss possible extensions of VisTA into sequence query and detection, as well as the potential for the visualizations to encourage self-directed learning in students.

In summary, our contributions are: i) a characterization of the challenges for teachers interested in using intelligent tutor data in their classroom planning, ii) the VisTA system, designed to visualize student interactions with the Apprentice Tutor intelligent tutoring system, and iii) a user evaluation with five teachers demonstrating how VisTA helps them better interpret intelligent tutor data, and gain insights into student problem-solving provenance.
Taken together, these insights from VisTA can help teachers better integrate tutors into their curriculum, and tailor their responses to individual student needs during classroom instruction.

\section{Related Works}

\subsection{Visualizing Provenance for Online Learning}
Many online learning systems, from Massive Open Online Courses (MOOCs) to learning management software (LMSs), increasingly deliver more sophisticated educational content to learners from diverse backgrounds \cite{singh2019many, moore2011learning}.
These systems collect rich data about student interactions and performance, which have in turn motivated the development of visual analytics tools that help teachers make sense of this data and gain insight into how students learn (see \cite{emmons2017mooc, dewan2021review, zhang2022towards} for recent surveys of visual analytics in online learning).
Of these, visualizations of learner trajectories have been widely adopted for understanding student activity and how students navigate through course content \cite{goulden2019ccvis, lundqvist2019visualising}.
Learner trajectories are paths by which learning takes place, and have been advocated as a useful means to guide student-centered teaching \cite{sztajn2012learning, wilson2015teachers} and optimize online course design \cite{ginda2019visualizing}.
However, most visualizations of learner trajectories have focused on when and how students access course content.
They do not account for the types of fine-grained interaction data collected by intelligent tutoring systems, which are more concerned with student problem solving provenance and skill mastery.

Prior work have found that provenance visualizations that depict the history of user actions can help us better understand user goals \cite{xu2020survey, dou2009recovering}, evaluate their performance \cite{battle2019characterizing, madanagopal2019analytic}, and assess their analytic strategies \cite{ragan2015characterizing, dou2009recovering}.
In particular, user interactions, defined as \textit{``the history of user actions and commands with a system''} \cite{ragan2015characterizing}, were a common type of data typically captured for provenance visualization \cite{xu2020survey, gomez2012modeling, brown2014finding, battle2019characterizing, stitz2018knowledgepearls}.
Although diverse visualization types have been proposed for provenance visualization in a range of application domains (for recent surveys, see \cite{xu2020survey} and \cite{battle2019characterizing}), there are, to the best of our knowledge, no existing visual analytics systems for exploring student problem solving provenance based on interaction data collected by intelligent tutoring systems.

\subsection{Visualizations for Intelligent Tutors}
Intelligent tutors are computer-based systems with two key models: (1) a knowledge-tracing model that estimates what a student knows, enabling personalized recommendations about which skills need more practice; and (2) an expert model that can provide learners with contextualized hints, worked examples, and correctness feedback \cite{vanlehn2006behavior}.
Benefits of these tutors have been widely acknowledged by researchers and educators over the years \cite{Pane1967,koedinger1997,Aleven2009}.
For instance, an intelligent tutor for algebra problem solving, PAT, was used in the Pittsburgh Urban Math Project (PUMP), and demonstrated one of the largest intelligent tutor deployments within a school system \cite{koedinger1997}.
These systems are also well liked by teachers because they provide individualized scaffolding and guidance to learners as they practice problem solving, allowing teachers to allocate time to other pedagogical tasks.

However, only a few intelligent tutors have made efforts to visualize student analytics as dashboards for teachers \cite{aleven2010developing, holstein2010luna, xhakaj2017effects, aleven2022dashboard} or students \cite{hou2022design}.
Findings from these prior works demonstrate that dashboards helped teachers develop their knowledge of individual students, and influenced their lesson plans \cite{xhakaj2017effects, aleven2022dashboard}, suggesting that visualizations of intelligent tutor data can help teachers better understand and respond to students’ needs during classroom instruction.
However, the majority of these dashboards only visualized aggregate data such as average skill mastery and average time spent on practice.
In cases where stepwise data was available, such as in the Tutti dashboard \cite{aleven2022dashboard}, each step could only be viewed as complete interface snapshots or replays, which can make it challenging for teachers to compare across students and summarize common patterns in problem-solving approaches.
Similarly, the Many Eyes Word Tree graphic proposed by \cite{shanabrook2012visualization} used stepwise data to depict sequential patterns of student states using a word tree diagram. Although this was helpful for algorithmic pattern discovery and prediction tasks, the graphic was not ultimately presented to students or teachers.

In this paper, we build on existing studies to develop visualizations of student interactions as they work through problems in the Apprentice Tutor platform.
We aim to provide insight into detailed student problem solving provenance and skill mastery.
In our evaluations, we validate that these visualizations can help teachers better integrate tutors into their curriculum, and tailor their responses to individual student needs during classroom instruction.

\section{Apprentice Tutors}
\label{sec:apprentice_tutors}
Since many different intelligent tutors have been developed to date, each catering to different topics, courses, and users, we decided to focus our study on a specific intelligent tutor, Apprentice Tutors\footnote{https://tutor.apprentice.ai}.
The system currently provides access to four mathematics \textbf{tutors}: Factoring Polynomials, Exponents, Radicals, and Rational Equations.
The system allows learners to engage with the content from any web browser, offering a convenient and flexible learning platform. The Apprentice Tutors uses Bayesian Knowledge Tracing (BKT) \cite{corbett1994knowledge}, a widely used knowledge tracing approach, to estimate students' mastery of each skill.
Using these skill estimates, the system adaptively selects problems for the student to practice next that will enhance the student's learning.


\begin{figure*}[!t]
  \includegraphics[width=\linewidth]{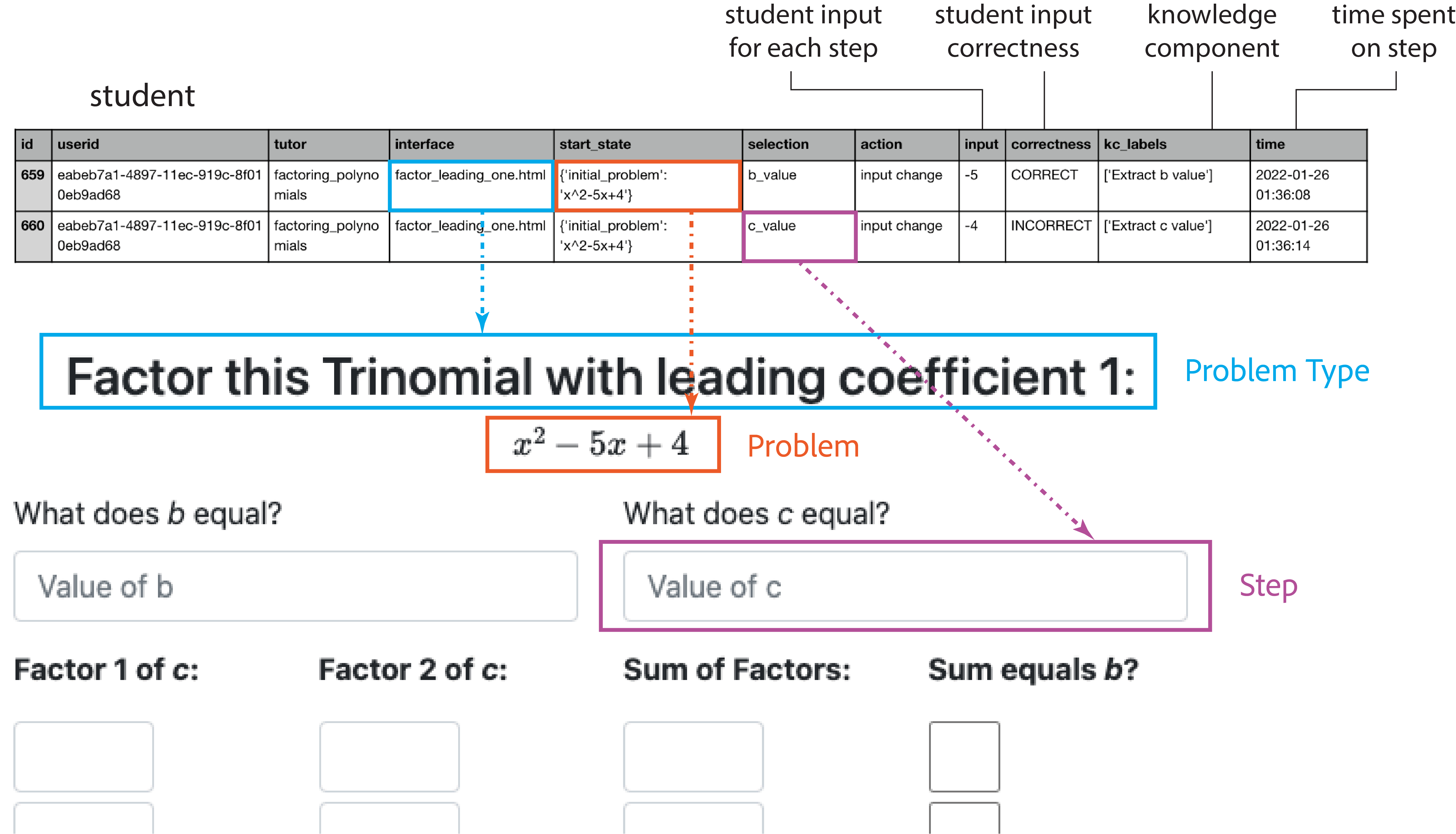}
  \caption{\textit{Top:} An example of two rows of data collected by Apprentice Tutors. \textit{Bottom:} How data columns map to different parts of the system. Each row records a single interaction by a student. From left to right, each interaction has 11 attributes: \textit{ID}, \textit{userID} (anonymized student ID), \textit{tutor}, \textit{interface} (problem type), \textit{start\_state} (problem), \textit{selection} (input box/step being attempted), \textit{action} (interaction type), \textit{input} (student input), \textit{correctness} (correctness of input), \textit{kc\_labels} (knowledge component/skill associated with this step), and \textit{time} (recorded time of student interaction).}
  \Description{Figure showing a sample problem on the Apprentice Tutors along with a table that stores data corresponding to each input field on the problem. The figure also highlights “Factor this Trinomial with leading coefficient 1” as the Problem Type, “x^2 -5x+4” as the Problem, and each input field as a “Step”. In the data table, “userId” corresponds to “student”, “input” corresponds to “student input for each step”, “correctness” corresponds to “student input correctness”, “kc_labels” corresponds to “knowledge component”, and “time” corresponds to “time spent on step”.}
  \label{fig:walkthrough}
\end{figure*}

Upon entry, students can select one of four tutors.
For each tutor, they can either opt for a specific \textbf{problem type} or an adaptive problem approach.
For example, the Factoring Polynomials tutor has multiple problem types including: ``leading coefficient of 1'', ``grouping method'', ``slip and slide method'', and ``area box method''.
The four problem types provide students with guidance on how to factor polynomials using various methods.
After selecting a problem type, a \textbf{problem} is dynamically generated for the student.

Adaptive problem selection is a core capability of the Apprentice Tutors that personalizes the student's learning experience by targeting skills that the student has yet to master.
Skills within the Apprentice Tutors are referred to as knowledge components (KCs), defined as acquired units of cognitive function that can be inferred from performance on a set of related tasks \cite{Koedinger2012}.
Each {\bf step} in a problem is associated with a skill or {\bf KC label}, and all problems of the same problem type have the same steps, which usually occur in the same sequence.
When students correctly complete a step associated with a KC, the system updates the associated student mastery estimate.
If students choose adaptive problem selection (instead of a single problem type), the tutor adaptively selects problem that exercise unmastered skills until the system estimates that the student has mastered all of the relevant KCs.

Figure \ref{fig:walkthrough} \textit{(Bottom)} illustrates a dynamically generated ``leading coefficient 1'' problem, accompanied by two rows of captured usage data (Fig. \ref{fig:walkthrough} \textit{Top}) that correspond to two consecutive interactions.
In this example, the problem is $x^{2} - 5x + 4$.
In the first two steps, students must practice skills (or KCs) for determining the $b$ and $c$ values from the polynomial.
In this example, the student has correctly \textbf{input} that $b= -5$, but incorrectly input that $c= -4$.
The system records the \textbf{time} spent and the \textbf{correctness} for each interaction.
If an interaction is incorrect, the student can make multiple attempts until their input for that step is correct.
They can also request hints by using the \textit{Ask for Hint!} button. Incorrect answers and hints decrease a student's estimated skill mastery.
The Apprentice tutors offer a three-tiered hint system: the first points to the required input box, the second rephrases the question, and the third provides the answer.
Students must provide the correct input for all steps, arrive at the final factored answer of $(x-4)(x-1)$, and click the \textit{Done} button to complete the problem.
A problem where all steps are correctly solved is considered ``complete'', otherwise it is considered ``incomplete'' (e.g. if the student stops halfway).
Hint requests for each step are also recorded.

\section{Design Study}
\label{sec:design_study}


To inform the design of the Apprentice Tutors visualizations, we conducted a formative focus group study session to understand how educators currently use intelligent tutors in the classroom, as well as how they would like to use and analyze intelligent tutor data.
We recruited five teachers (P1-5, 4F/1M) from a local state university\footnote{Location not affiliated with authors.}
who have deployed the Apprentice Tutors in their classrooms for at least a semester prior to our focus group.
Participants were compensated with a \$25 gift card for their time.
The sharing of anonymous student data and the details of the focus group protocol were approved by our Institutional Review Board (IRB) prior to the study.
Participants also provided their documented consent before the session.

The focus group took place virtually, and all participants were present at the same session.
Since none of the teachers had accessed the Apprentice Tutor data prior to the focus group, we first introduced them to the data set (Fig. \ref{fig:walkthrough}).
The data was anonymized beforehand, and no identifiable information about students was shared.
The session was structured around the following questions:
\begin{itemize}[align=parleft, left=0pt..8pt]
    \item[-] Do you have any questions about the data?
    \item[-] Is there anything in the data that is more interesting/useful?
    \item[-] Is there anything in the data that is not interesting/useful?
    \item[-] Is there anything else you want to see in the student data?
    \item[-] Outside of Apprentice Tutors, do you typically collect and analyze data about your students?
    \item[-] (If yes:) What are your processes and tools?
\end{itemize}
We asked follow-up questions where appropriate.
The formative study lasted for about an hour.




\subsection{Design Challenges}
\label{sec:design_challenges}
Based on findings from our focus group, we used an affinity diagramming process to identify main themes in user needs and requirements.
This was supplemented by findings from prior works \cite{aleven2022dashboard} to derive three design challenges that should be addressed in our visual analytics system.

\medskip
\noindent\textbf{C1  }
\textbf{Summarizing student engagement.}
Our participants described the need for an intuitive summary of student engagement in terms of which problems were being attempted and aggregate performance metrics.
As P4 described, \textit{``Overall, I need to know how many [problems] students answered correctly and how many incorrect.''}
P2 also mentioned that she wanted to know the names of students who use the tutor, and the frequency of their participation.
Our interface should allow teachers to quickly identify which of their students are engaging with the intelligent tutors and rapidly evaluate student accuracy for attempted problems.


\medskip
\noindent\textbf{C2  }
\textbf{Identifying specific issues at each problem-solving step.}
Participants wanted to understand how well their students understood each knowledge component or problem step.
For P2, her current practice at the end of each topic was to ask students to \textit{``go back and look at it and see where you guys get lost''} and provide feedback for places where they need revision.
Overall, teachers wanted to identify and focus on common steps their students found challenging.
Thus, our visualizations should help teachers understand step-wise problem-solving provenance across all students in order to identify challenging steps that need further attention.


\medskip
\noindent\textbf{C3  }
\textbf{Comparing students' performance over time.}
Finally, participants also wanted to look at performance data for students in context of time spent. 
Teachers mentioned that they were currently using activity logs to understand \textit{``how hard [students] are working and how much time they're dedicating''} (P3).
They also found that there was \textit{``a correlation between how much time [students] are spending on practicing the homework and their grade''} (P3).
As such, our system should visualize both total time spent as well as detailed temporal break downs of each step.

\section{The V\MakeLowercase{is}TA System}
Based on the challenges identified from our focus group session, we designed the VisTA (Visualizations for Tutor Analytics) system.
VisTA is an analytics tab that teachers can access through the Apprentice Tutors platform.
All visualizations are implemented in JavaScript and D3 \cite{bostock2011d3}.
In the following sections, we describe the main views in VisTA and walk through example use cases.
When VisTA is deployed, teachers using VisTA will be able to see student names. However, we omit names and any other identifiable information in screenshots throughout the paper to preserve anonymity.

\subsection{System Implementation}

\begin{figure*}[h]
  \includegraphics[width=\textwidth]{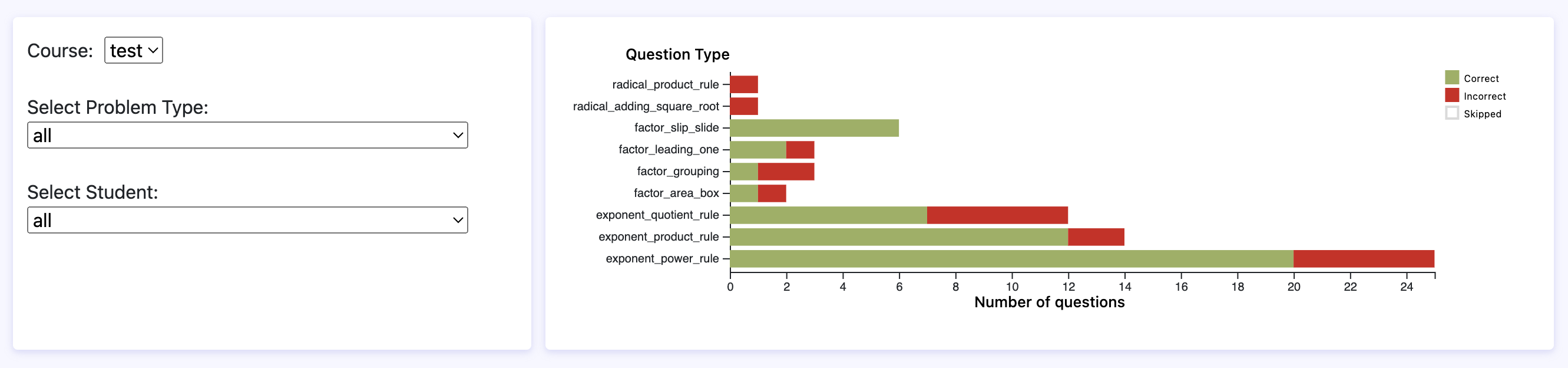}
  \caption{The \textit{Overview} visualization showing proportion of correct \textcolor{correct}{(green)}, incorrect \textcolor{incorrect}{(red)}, and skipped \textcolor{gray}{(gray)} questions attempted by students for each problem type. Users can click on the respective labels in the legend to select and deselect the groups they would like to include in the histogram. The control panel on the left can be used to select a particular problem type or student to visualize.}
  \Description{Figure showing an Overview visualization using a stacked horizontal bar chart. Y-axis of the chart plots the Problem Types, and the x-axis plots the Count Attempts. The bars are colored green for Correct steps attempted, red for Incorrect steps attempted and grey for Skipped steps attempted.}
  \label{fig:overview}
\end{figure*}

\subsubsection{Overview.} When teachers access VisTA, they will first see a horizontal stacked bar chart of attempted problems broken down by problem type (Fig. \ref{fig:overview}).
This histogram provides an \textit{overview of student participation} \textbf{(C1)}, and allows teachers to quickly identify problem types that students have practiced more.
Each bar in the histogram is further divided into segments representing the proportion of \textit{practice questions that were \textcolor{correct}{completed correctly} versus those that were \textcolor{incorrect}{incorrect or partially complete}} \textbf{(C1)}.
There is a third group of \textcolor{gray}{skipped} questions that is deselected by default.
Skipped questions are those that were started but which the students did not interact with (i.e. the student did not attempt any steps of the question).
teachers can click on the respective labels in the legend to include skipped questions in the overview histogram.
From Fig. \ref{fig:overview}, we can see that problems about exponents (\verb|exponent_product_rule| and \verb|exponent_quotient_rule|) were the most attempted problem types and students completed them accurately.
Comparatively, students tended to be more inaccurate on problems about radicals.

\begin{figure}[!t]
  \includegraphics[width=\linewidth]{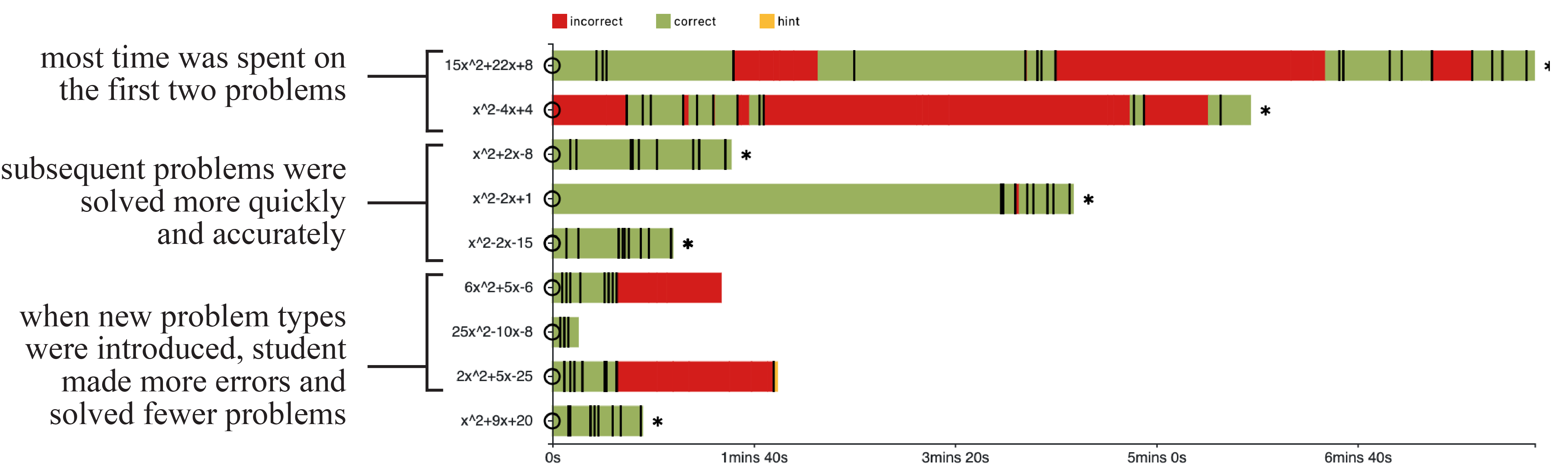}
  \caption{The \textit{Student view} showing all problems of a particular problem type attempted by a single student. We can see that the student spent the most time and made the most \textcolor{incorrect}{mistakes} on the first two problems, but solved subsequent problems quickly and more accurately. When new problem types with new knowledge components (solving polynomials with first coefficient $>$1) were introduced, the student started to make more mistakes and left the problems incomplete.}
  \Description{Visualization showing all problems of a particular problem type attempted by a single student.}
  \Description{Figure showing a Student view using a horizontal stacked bar chart. Y-axis of the chart plots the Problem, and the x-axis plots the Time. The bars are colored green for Correct steps attempted, red for Incorrect steps attempted and yellow for Hints used on any step. A black vertical line at the end of a bar indicates the end of an attempted step and the beginning of the next. A small black circle at the beginning of the bars on the y-axis indicate that the student has started the problem, and an asterisk at the end of each stacked bar indicates that the student has completed the problem. The first 2 stacked bars are the longest indicating that those 2 problems took the longest time. The subsequent 3 problems have all green stacked bars (with minimal red bars) indicating that the student solved them quickly and accurately. The following 3 problems introduced more knowledge components and their stacked bars had more red bars indicating that the student’s error rate increased. The intelligent tutor reduced the difficulty with the last problem and their stacked bars are all green and short in length indicating that the student solved the problem quickly and accurately once again.}
  \label{fig:student}
\end{figure}

\subsubsection{Student View.}
From the drop-down menu on the left, teachers can select a particular student in their class to view \textit{all problems attempted by that student} \textbf{(C1)} in a single horizontal stacked bar chart (Fig. \ref{fig:student}).
Each problem is divided into its component steps by black lines, and each step is further divided into separate interactions.
We use the same color scheme as the feedback provided in the tutor interface (Section \ref{sec:apprentice_tutors}), where \textit{\textcolor{correct}{green segments} represent a correct response, \textcolor{incorrect}{red segments} represent an incorrect response, while \textcolor{hint}{yellow segments} indicate when students asked for a hint} \textbf{(C2)}.
The same color palette is used for all visualizations in VisTA, and was selected to be colorblind safe.
Hovering over a segment shows a tooltip with details of the interaction, such as \textit{the KC associated with the step and student input value} \textbf{(C2)}.
An asterisk (*) indicates that \textit{the student completed a problem correctly, otherwise, the problem was incomplete or not attempted} \textbf{(C2)}.
In this visualization, the width of each segment represents \textit{the duration of each interaction}, while the total width of all segments reflects \textit{total time spent by the student on the problem} \textbf{(C3)}.

\begin{figure*}[!t]
  \includegraphics[width=\textwidth]{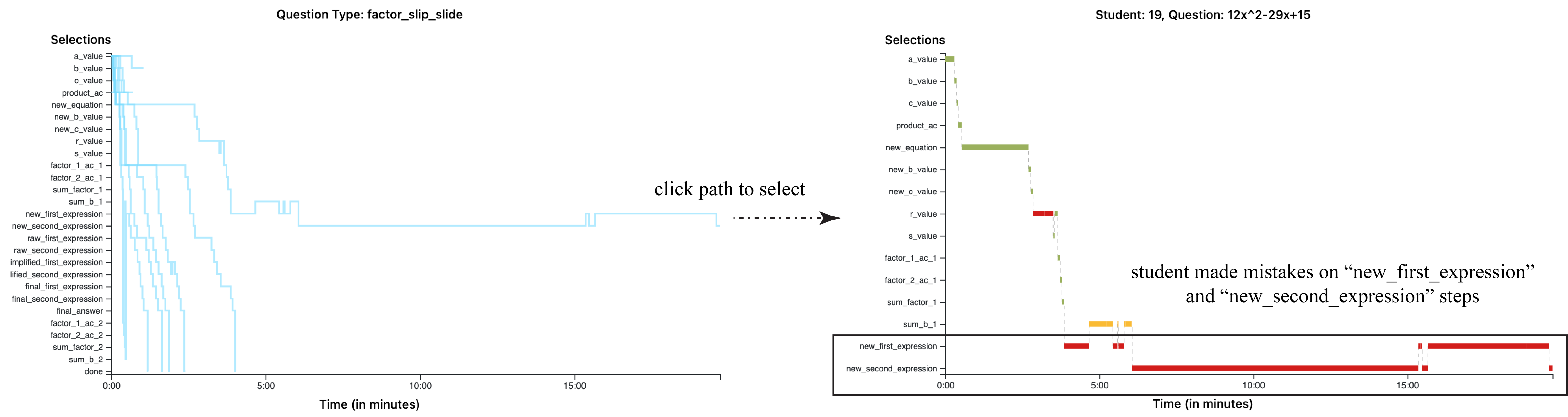}
  \caption{\textit{Left:} The \textit{Problem Type view} uses a step line chart to depict all student attempts of a particular problem type. In this example, we see that while most attempts were completed in under 5 minutes, there was one student attempt that took more than 15 minutes and was left incomplete. \textit{Right:} Clicking on a path brings up the \textit{Details view}, which shows the breakdown of each step. Here, the student appears to have made multiple mistakes on the ``new\_first\_expression'' and ``new\_second\_expression'' steps before leaving the problem incomplete.}
  \Description{Step line chart showing all student attempts of a particular problem type. Time (in minutes) is visualized on the x-axis. Steps of the problem are visualized on the y-axis.}
  \label{fig:problem}
\end{figure*}

\subsubsection{Problem Type view.}
From the same drop-down menu on the left, teachers can also select a particular problem type to analyze.
For each problem type, we visualize all student attempts in a single step line chart (Fig. \ref{fig:problem}, \textit{Left}).
The steps of the problem are listed along the \textit{y}-axis from top to bottom.
Each student attempt is represented by a single blue line.
\textit{The amount of time spent on a step} \textbf{(C3)} is visualized along the \textit{x}-axis, and corresponds to the length of the horizontal line segment.
As a student works through a problem, we expect to see their line descend from the top left corner to the bottom right corner, \textit{demonstrating progression through the problem} \textbf{(C1)}.
Lines that terminate in the middle of the chart indicate an incomplete attempt.


\subsubsection{Details view.}
Finally, teachers can click on a single path in the \textit{Problem Type view} or use the drop-down menu to see the details of \textit{how a single student attempted a single practice problem} (Fig. \ref{fig:problem}, \textit{Right}).
This view is a combination of the stacked bar chart from the \textit{Student view} and the step line chart from the \textit{Problem Type view} for a single problem.
In this visualization, the steps for the problem are listed on the \textit{y}-axis and the time spent is represented on the \textit{x}-axis.
As a student works through a problem, the line descends from the top left corner to the bottom right corner, with each step broken down further into detailed interactions and correctness.
Taken together, this \textit{Details view} provides information on accuracy, time, and interaction sequence \textbf{(C2, C3)}, allowing teachers to not just evaluate how well students can solve a problem, but also identify where mistakes occur.

\subsection{Use Cases}
In the previous sections, we have described the visualizations in VisTA and how they address the challenges surfaced during our focus group session.
We now walk through three use cases demonstrating the insights that teachers might gain from using VisTA.

\subsubsection{Understanding student engagement.}
Ana is a math teacher who has made the Apprentice Tutors available to her students who want extra practice on the topics covered in class.
She is now interested in whether the tutors have been helpful to her students, so she navigates to the Analytics tab of the Apprentice Tutors platform, and begins exploring the visualizations available.
From the \textit{Overview} (Fig. \ref{fig:overview}), she sees that her students have practiced a large number of problems about exponents.
Comparatively, problems about factoring and radicals received less practice, and there was a greater proportion of inaccurate responses for those tutors.

\subsubsection{Understanding skill mastery from practice.}
Recalling that there is a student in her class who has been doing well on factoring, Ana further investigates whether the student has been practicing with the Apprentice Tutors.
She selects their name from the drop-down menu and, from the \textit{Student view}, finds that the student has indeed practiced a large number of problems related to factoring (Fig. \ref{fig:student}).
She notices that the student had struggled on the first two questions, making multiple inaccurate inputs.
However, they appeared to learn from the practice and solved three subsequent problems quickly and accurately.
Ana then notices that while the student was able to complete all factoring problems when the first coefficient was 1, as the Apprentice Tutors system started introducing problem types with new KCs (polynomials with first coefficient $>$1), those problems were left incomplete.
Ana makes a note to follow up with the student to see if they need help practicing skills to solve such problems.
She then decides to look at how other students have performed on the same problem type to see if this is a common challenge she needs to address in her upcoming class.

\begin{figure}[!t]
  \includegraphics[width=\linewidth]{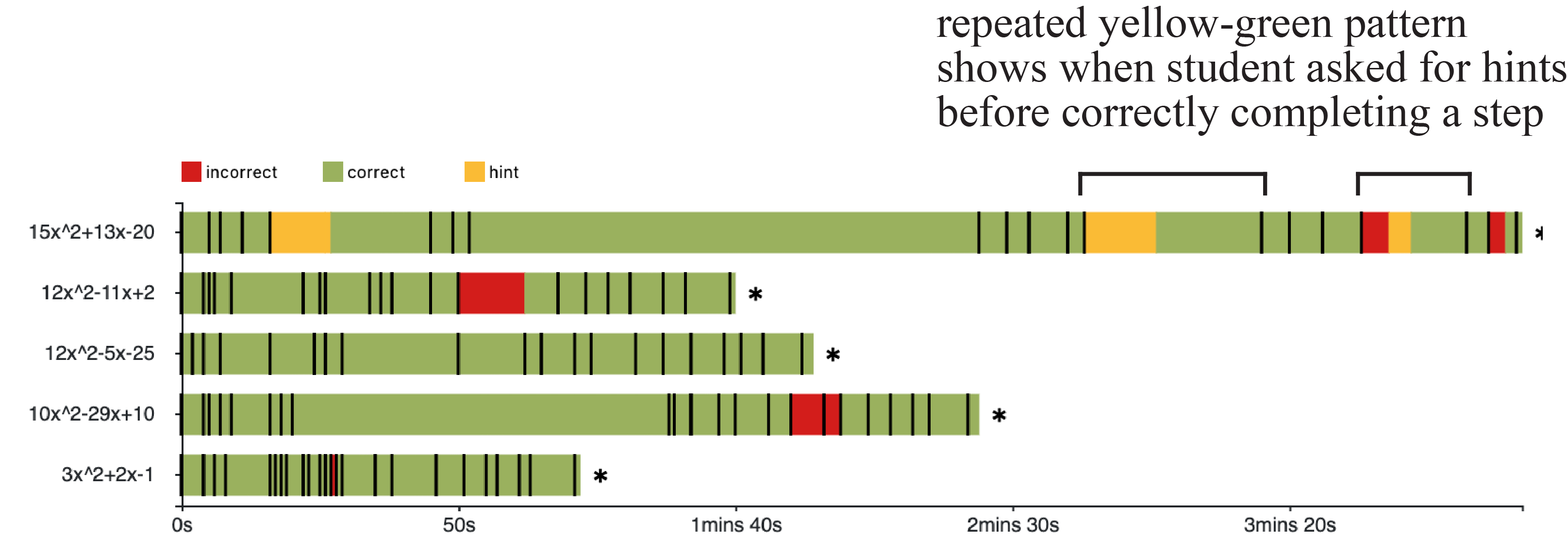}
  \caption{\textit{Student view} for a student whose interaction patterns show that they frequently asked for \textcolor{hint}{hints} while working through the problems.}
  \Description{Figure showing a Student view using a horizontal stacked bar chart. There are repeated yellow-green patterns for several steps in the problems indicating when the student asked for hints while completing the steps.}
  \label{fig:strategy}
\end{figure}

\subsubsection{Comparing usage patterns and problem solving strategies.}
When looking at the detailed interactions of another student in her class using the \textit{Student view}, Ana comes across an interesting strategy (Fig. \ref{fig:strategy}).
She realizes that this student has been using multiple hints to help them solve problems.
Majority of the hint usage is concentrated in the first problem, and as the student gained familiarity with the problem type, the number of hints requested also decreased.
She then contrasts this visualization with the student from Fig. \ref{fig:student}.
She realizes that the previous student rarely used hints, even when hints might have helped them correct inaccuracies and progress with the problem (Fig. \ref{fig:student}).
Ana makes a note to remind students of the hint feature.
She also wonders if the tutors can be improved by automatically prompting students to request hints based on detected student behavior.


\section{Evaluation}

We conducted a second focus group session and presented VisTA back to the same group of five teachers (see Section \ref{sec:design_study}) for feedback and evaluation.
Participants were compensated with a \$25 gift card for their time.
The sharing of anonymous student data and the details of the focus group protocol were approved by our Institutional Review Board (IRB) prior to the study.
Participants also provided their documented consent before the session.

During the session, we first demonstrated VisTA and all visualization views to the teachers.
All student names were omitted from the demonstration to preserve student privacy.
The subsequent discussion was broadly structured around the following questions:
\begin{itemize}[align=parleft, left=0pt..8pt]
    \item[-] Is there anything in the visualization prototypes that is more interesting/useful for you?
    \item[-] Is there anything in the visualization prototypes that is not interesting/useful for you?
    \item[-] Is there something you would like to know that is currently not shown in the visualizations?
\end{itemize}

During the session, we followed up on the feedback provided, particularly comments about how teachers might use the visualizations, any visual patterns they found interesting, and areas for improvement.
Finally, teachers were invited to ask any questions they had of us.
The evaluation focus group lasted around an hour.


\subsection{Results}

Overall, participants found VisTA to be useful.
We used an affinity diagramming process to categorize their feedback into two major themes that emerged during the focus group: tailoring responses to student performance, and identifying steps in a problem that require more classroom instruction.

\subsubsection{Visualizing provenance helped supplement summary metrics and motivated nuanced responses to student performance.} \label{provenance}
Teachers in our study found particular value in being able to see the details of the student problem-solving provenance (\textit{Student view}, \textit{Problem Type view} and \textit{Details view}) combined with summary statistics about the overall accuracy (\textit{Overview}).
By breaking down each problem into individual steps and interactions, it was \textit{``useful to see what part of the process that they get tripped on''} (P3).
It also helped teachers identify if and when a student needed more assistance.
As P1 described, if it was just one step with an inaccurate input, \textit{``it might be a goofy mistake... but if I see red and red and red, then I need to contact the student and send them to tutors... we need to figure out what's going on''}.
Similarly, P3 mentioned that a single inaccurate response may not mean that \textit{``they're lacking the skills to complete the process, maybe it's that they don't know how to [use the intelligent tutor]''}.
In such cases, a student may make mistakes on problems in the beginning, but show improvements in their performance on subsequent questions as they gained familiarity with the tutor interface (such as in Fig. \ref{fig:student}).
As such, by inspecting the stepwise provenance of how a student solved problems, teachers can gain a more nuanced interpretation of when and why students make mistakes, and tailor their response to different students appropriately.
As P2 summarized: \textit{``I think it's good to see all the students together, but it's also good to see one student and how they do all over on all the tutors that they use''}.


\subsubsection{Aggregating and comparing solution paths helped highlight steps (KCs) that require more instruction.}
Teachers in our evaluation study found the \textit{Problem Type view} to be particularly useful.
By visualizing the details of multiple student attempts on the same chart, the system helped call attention to the steps and KCs that need more instruction.
Teachers can see \textit{``how many students are getting the middle part [and] which part I need to focus on it more''} (P2).
This helped them review their teaching and proactively consider how they might address the common challenges their students face.
In some cases, it might also prompt reflections on the design of the intelligent tutor and how it might be improved.
As P5 described, her insights after viewing the visualizations were such that \textit{``if several students are getting hints or missing in the same place, I would go back and look at the tutor and see if it's something in the tutor, or is it that they don't understand the terminology, or do I need to reteach that area. It makes me think all three of those.''}
Taken together, we see how the \textit{Problem Type view}, in particular, provides valuable insight into how students interact with the intelligent tutor, helping teachers reflect on both tutor design and individual student needs.

\section{Discussion}

In addition to providing feedback on the usability of the VisTA system, the teachers in our study also raised interesting avenues of future work, where VisTA might be extended to support new features and user groups.

\subsection{Temporal Sequence Query and Detection}
During the evaluation study, teachers agreed that not all patterns or errors in the visualization were equally important.
For example, a single inaccurate input could be an accident, but other patterns such as a sequence of \textcolor{incorrect}{red}-\textcolor{incorrect}{red}-\textcolor{incorrect}{red} inputs would indicate that they need to follow-up with the student (see Section \ref{provenance}).
Furthermore, a teacher who sees a concerning pattern of interactions may find it helpful to identify other students who also display this sequence.
This suggests one promising direction of future work into the automatic highlighting of interesting data-facts and salient event sequences.
Teachers can input user-defined temporal sequence queries to search for patterns of interest that are then brought to their attention.
For instance, while one teacher may be concerned about inaccuracies, another may be more interested in how their students have been using the hint feature.
This user-driven query function can complement manual data exploration, helping to surface insights specific to different user contexts.
Existing work into temporal sequence analysis \cite{guo2018visual, guo2017eventthread, gotz2014decisionflow, wongsuphasawat2012exploring, zgraggen2015s, lan2013temporal} and automatic data-fact generation \cite{srinivasan2018augmenting, law2020characterizing, chen2010click2annotate, brath2021automated, demiralp2017foresight} can provide guidance for how VisTA might be extended to automatically detect and visualize teacher-defined data facts.
In future work, we plan to build on these existing studies to implement a temporal sequence query feature to help teachers define and automatically highlight interesting patterns in intelligent tutor data.

\subsection{Supporting Self-Directed Learning}
Multiple teachers in our evaluation study also raised the potential for VisTA to be useful for their students as well.
P2 mentioned that the visualizations might be \textit{``more useful for the students than [for] the teachers''} because the visualizations could help students identify where they went wrong.
Seeing their accuracy for each step of a question might also lead students to reflect that \textit{``maybe I was on the right track, but I'm missing some part of this''} (P3) or that \textit{``they made a mistake [and] have to review that part before the test''} (P1).
This feedback is supported by theories of self-directed learning where learners are encouraged to set their own goals, monitor their performance, and seek out feedback \cite{robinson2020developing, garrison1997self}.
Prior work implementing such visualizations had demonstrated that showing students visualizations of their own performance prompted them to be more reflective of their behavior and reduced attempts to game the system \cite{hou2022design, xia2020using}.
These studies also found that visualizing step-wise temporal information and peer-related data were important factors that influenced self-reflection.
This suggests that many of the metrics and views in VisTA may be similarly effective for supporting self-directed learning in students.
Future studies can further explore this by expanding visualization design and development to include students using the Apprentice Tutors as well.

\subsection{Limitations and Future Deployment}

Due to scheduling constraints and teaching commitments, we tried to be mindful of the teachers' time when planning for the design study and conducted the requirements gathering and evaluations through two focus-group sessions instead of individual interviews.
Furthermore, since the same group of teachers participated in both sessions, it is likely that VisTA may not generalize well to other teachers and courses.
To address these concerns and to ensure that teachers obtain hands-on experience with VisTA instead of simply viewing a demonstration, we have currently deployed VisTA to all teachers at our partner institution.
Based on this long-term deployment, we plan to conduct a more in-depth evaluation of how, and to what extent, the VisTA system helps teachers adjust their classroom instruction, and whether this results in measurable outcomes in student classroom performance.

\section{Conclusion}

In this paper, we have provided a characterization of the challenges for teachers interested in integrating intelligent tutor data in their teaching.
Based on these challenges, we designed the VisTA system to visualize logs of student interactions with the Apprentice Tutor intelligent tutoring system.
VisTA is, to the best of our knowledge, the first analytics tool developed to visualize student provenance data from intelligent tutors.
In a user evaluation with five teachers, we demonstrated how VisTA helps them better interpret intelligent tutor data, gain insights into student problem-solving provenance, and tailor their follow-up responses to student needs during classroom instruction.
We have currently implemented VisTA on the Apprentice Tutors platform.
In future work, we plan to look at how VisTA is used in a long-term deployment over the course of a semester.
We also plan to examine how VisTA might be extended to support context-dependent sequence detection for teachers and self-directed learning in students.

\begin{acks}
The authors would like to thank the Georgia Tech Visualization Lab for their feedback and suggestions, as well as Jacob Dallas-Main, Adie Shimandle, and the teachers of Chattahoochee Technical College for organizing and attending the focus group sessions.
This work is funded in part by NSF 2112532 and the IBM PhD Fellowship.
\end{acks}

\bibliographystyle{ACM-Reference-Format}
\bibliography{sample-base}


\end{document}